\documentclass[letterpaper, 10 pt, journal, twoside]{ieeetran}
\usepackage{amsmath}
\usepackage{cite}
\usepackage{amssymb}
\usepackage[ruled,linesnumbered,noline]{algorithm2e}
\usepackage{float}
\usepackage{textcomp}
\newtheorem{thm}{Theorem}

\newtheorem{definition}{Definition}
\newtheorem{problem}{Problem}
\newtheorem{assum}{Assumption}

\usepackage{graphicx}
\usepackage{tikz}
\usepackage{float}
\usetikzlibrary{arrows}
\usetikzlibrary{arrows.meta}
\usetikzlibrary{arrows,positioning} 
\usetikzlibrary{shapes}
\usetikzlibrary{calc}
\usetikzlibrary{fadings}
\usetikzlibrary{decorations.pathreplacing}
\usetikzlibrary{decorations.text}
\newcommand*{\QED}[1][$\blacksquare$]{%
\leavevmode\unskip\penalty9999 \hbox{}\nobreak\hfill
    \quad\hbox{#1}%
}
\usepackage{amsfonts}

\def\next{\bigcirc}
\def\eventually{\Diamond}
\def\until{\mathcal{U}}
\def\true{\top}

\tikzfading[name=fade right,
left color=transparent!0, right color=transparent!100]
\tikzfading[name=fade out, 
    inner color=transparent!0,
    outer color=transparent!100]
\definecolor{dcompb}{RGB}{157,35,0} 
\definecolor{mblue}{RGB}{0.176, 0.380, 0.659}
\definecolor{dgreen}{RGB}{15,111,3}
\definecolor{dred}{RGB}{.65,.176,0}
\definecolor{dblue}{RGB}{.098,.243,.424}
\usepackage[top=57pt, left=48pt, right=48pt, bottom=43pt]{geometry}
\begin{document}

\title{Safe Learning for Uncertainty-Aware Planning via Interval MDP Abstraction\vspace*{3pt}}


\author{Jesse~Jiang,~\IEEEmembership{Student Member,~IEEE,} Ye~Zhao,~\IEEEmembership{Member,~IEEE,} and Samuel~Coogan,~\IEEEmembership{Senior Member,~IEEE}
\thanks{This work was supported in part by the National Science Foundation under grant \#1924978.}
\thanks{Jesse Jiang and Samuel Coogan are with the School of Electrical and Computer Engineering, Georgia Institute of Technology, Atlanta, GA 30332 USA (e-mail: jjiang@gatech.edu, sam.coogan@gatech.edu). S. Coogan is also with the School of Civil and Environmental Engineering. }
\thanks{Ye Zhao is with the School of Mechanical Engineering, Georgia Institute of Technology, Atlanta, GA 30332 USA (e-mail: ye.zhao@me.gatech.edu).}}
%

\maketitle
\thispagestyle{empty}
\begin{abstract}
We study the problem of refining satisfiability bounds for partially-known stochastic systems against planning specifications defined using syntactically co-safe Linear Temporal Logic (scLTL). 
We propose an abstraction-based approach that iteratively generates high-confidence Interval Markov Decision Process (IMDP) abstractions of the system from high-confidence bounds on the unknown component of the dynamics obtained via Gaussian process regression. In particular, we develop a synthesis strategy to sample the unknown dynamics by finding paths which avoid specification-violating states using a product IMDP. We further provide a heuristic to choose among various candidate paths to maximize the information gain. Finally, we propose an iterative algorithm to synthesize a satisfying control policy for the product IMDP system. We demonstrate our work with a case study on mobile robot navigation.
\end{abstract}


\section{Introduction}
\IEEEPARstart{A}{bstraction}-based approaches for verification and synthesis of dynamical systems offer computationally tractable methods for accommodating complex specifications \cite{Tabuada2009}. In particular, Interval Markov Decision Processes (IMDP) \cite{belta_formal_2017}, which allow for an interval of transition probabilities, provide a rich abstraction model for stochastic systems. As compared to stochastic control\cite{Mark_2020}, abstraction-based methods allow for more complex specifications to be considered and have been widely used for hybrid stochastic systems \cite{Lavaei2021}.

The transition probability intervals in IMDP abstractions have typically modeled the uncertainty which arises from abstracting the dynamics of continuous states in discrete regions \cite{baier_model-checking_2003}. However, partially-known stochastic systems, which show promise for modeling a wide range of real-world systems, add unknown dynamics which contribute further uncertainty. Previous works model this uncertainty by assuming that some prior data on the dynamics are available \cite{ahmadi-safety-2017}.

The paper \cite{jackson2021strategy} is the first to address the problem of modeling unknown dynamics in stochastic hybrid systems via the use of IMDP abstraction in combination with \textit{Gaussian process} (GP) regression \cite{Williams96gaussianprocesses}. GP regression can approximate unknown functions with arbitrary accuracy and also provides bounds on the approximation uncertainty \cite{akametalu_reachability-based_2014}.

The main contribution of this work is to develop a method for sampling the unknown dynamics of a stochastic system online in order to reduce abstraction error and increase the probability of satisfying a syntactically co-safe linear temporal logic (scLTL) specification \cite{belta_formal_2017}.

Our goal is to find a control policy which guarantees the satisfaction of a scLTL specification with sufficient probability. However, we assume a stochastic noise which creates unavoidable perturbation. The system also has unknown dynamics which we estimate with Gaussian processes. This creates an estimation error which increases uncertainty in state transitions and which we aim to reduce by sampling the unknown dynamics. Thus, this paper focuses on the problem of safe learning to allow online exploration rather than a static planning problem using previously collected data samples as in \cite{jackson2021strategy}.

Our approach is as follows. First, we estimate the unknown dynamics of the system using Gaussian processes and construct a high-confidence IMDP abstraction. We then develop an algorithm for finding \textit{nonviolating cycles} in a product IMDP of the system abstraction combined with a finite automaton of the scLTL specification which allow the dynamics of the system to be sampled without violating the specification. We develop a heuristic for evaluating candidate cycles in order to maximize the uncertainty reduction gained from sampling. Finally, we propose an iterative method to sample the state-space, thereby decreasing the uncertainty of a GP estimation of the unknown dynamics until a satisfying control policy for the system can be synthesized or a terminating condition such as a maximum number of iterations has been reached. We utilize sparse GPs \cite{leibfried2021tutorial} to improve computational efficiency. We demonstrate our method on a case study of robotic motion planning. 

\section{Problem Setup}\label{section:Setup}
Consider a discrete-time, partially-known system
\begin{equation}\label{System Dynamics}
x[k+1] = f(x[k]) + u[k] + g(x[k]) + \nu[k]
\end{equation}
where $x \in X \subseteq \mathbb{R}^n$ is the system state, $u \in \mathbb{R}^n$ is the control action, $f(x)$ is the known dynamics, $g(x)$ is the unknown dynamics to be learned via GP regression, $\nu$ is stochastic noise, and time is indexed with brackets. This system has applications in, \textit{e.g.}, biology \cite{Julius2008}, communications \cite{ALTMAN19991937}, and robotics \cite{shamsah_integrated_2021}.
\begin{assum} 
\label{assum:1}
1) Each dimension $\nu_i[k], i=1,\ldots,n$ of $\nu$, is an independent, zero mean random variable with stationary, symmetric, and unimodal distribution $\rho_{\nu_i}$ and is $\sigma_{\nu_i}$-sub-Gaussian, \emph{i.e.}, the distribution tail decays at least as fast as a Gaussian random variable with variance $\sigma_{\nu_i}^2$.

2) Given a data set $D=\{(z^j,y^j)\}_{j=1}^m$ where $y^j$ is an observation of $g(z^j)$ perturbed by $\sigma_{\nu_i}$-sub-Gaussian noise, 
it is possible to construct an estimate $\hat{g}^D(x)$ of $g$ and bound the estimation error between $g(x)$  and $\hat{g}^D(x)$ by
some high-confidence bound $\gamma^D(x)$. Thus, \begin{equation}\label{eq:GPbound}
    g_-^D(x)=\hat{g}^D(x)-\gamma^D(x), \quad g_+^D(x)=\hat{g}^D(x)+\gamma^D(x)
\end{equation}
are high-confidence bounds on $g$, \emph{i.e.}, $g^D_-(x)\leq g(x)\leq g^D_+(x)$ with high confidence. For simplicity, we drop the superscript $D$ when the dataset is clear.
\end{assum}


\begin{assum}
\label{assum:partitions}
The state-space $X$ is bounded and is partitioned into hyper-rectangular regions $\{X_q\}_{q\in Q}$ defined as
\begin{equation}
\label{eq:part}
    X_q=\{x\mid \ a_q\leq x\leq b_q\}\subset X,
\end{equation}
where the inequality is taken elementwise for lower and upper bounds $a_q,b_q\in\mathbb{R}^n$ and $Q$ is a finite index set of the regions. Each region has a center 
$c_{q}={(a_q+b_q)}/{2}$. Additionally, the system possesses a labeling function $L$ which maps hyper-rectangular regions to observations $O$.

\end{assum}

Define feedback controllers $K_{{q}}(\cdot\ ;\hat{g}):X\xrightarrow{}X$ 
as 
\begin{equation}
    \label{eq:feedback}
    K_{q}(x;\hat{g})=c_{q}-f(x)-\hat{g}(x).
\end{equation}

\begin{figure}[t]
    \centering
    
\begin{tikzpicture}[xscale=3.3, yscale=1.6]
  \draw[line width=1pt] (0,0) grid (2,2);
    \draw[fill=dred!60] (.2,1.3) rectangle (.8,1.7);
 \fill[dgreen, path fading=fade out]
 (.4,1.2) rectangle (1.2,1.8);
 \fill[dgreen, path fading=fade out]
 (-0.2,1.4) rectangle (0.6,2);
 \node[fill=dblue,inner sep=2pt, circle,label=right:{$x[k]$}] (a) at (.2,.4) {};
 \node[fill=dblue,inner sep=2pt, circle,label=right:{$x[k+1]$}]  (b) at (1.1,1.4) {};
 \draw[->] (a) to[bend left] (b);
  \node[fill=yellow, draw=black,inner sep=2pt, circle, label={[label distance=-4pt]left:{$c_{2}$}}] at (.5,1.5) {};
\node[fill=dgreen,draw=black,inner sep=2pt, circle] at (.8,1.5) {};
\node[fill=dred,draw=black,inner sep=2pt, circle] at (.2,1.7) {};
  \node at (.1,.1) {$X_1$};
  \node at (1.9,.1) {$X_4$};
  \node at (.1,1.9) {$X_2$};
    \node at (1.9,1.9) {$X_3$};

\coordinate (legstart) at (-.1,-.4);
    \draw[fill=dred!60] (legstart) rectangle +(.2,.2);
    \node[align=left, rectangle, minimum width=3in,text width=3in] at  ($(legstart)+(1.4,.1)$) {\footnotesize $=$ uncertainty from high-confidence bounds on $g(x)$};

 \fill[dgreen, path fading=fade out] ($(legstart)+(0,-.2)$) rectangle +(.2,.2);
        \node[align=left, rectangle, minimum width=3in,text width=3in] at  ($(legstart)+(1.4,-.1)$) {\footnotesize$=$ uncertainty from stochastic noise $\nu$};
        \node[fill=dgreen,draw=black,inner sep=2pt, circle] at ($(legstart)+(.1,-.3)$){};
                \node[align=left, rectangle, minimum width=3in,text width=3in] at  ($(legstart)+(1.4,-.3)$) {\footnotesize$=x_{\max}$};
        \node[fill=dred,draw=black,inner sep=2pt, circle] at ($(legstart)+(1.1,-.3)$){};
                \node[align=left, rectangle, minimum width=3in,text width=3in] at  ($(legstart)+(2.4,-.3)$) {\footnotesize$=x_{\min}$};
\draw[->,line width=1pt] (0,0) -- node[below,pos=1]{$x_1$} (2.2,0);   \draw[->,line width=1pt] (0,0) -- node[left,pos=1]{$x_2$} (0,2.2);               
\end{tikzpicture}
    \caption{Feedback controller and calculation of transition probabilities. The controller targets the center of state $X_2$. The uncertainty in $\hat{g}(x)$ creates a nondeterministic region of transition (brown rectangle). The maximum probability of transitioning to state $X_3$ is found by centering stochastic noise at the point $x_{\max}$ closest to state $X_3$ (green point) and calculating the probability that the noise reaches state $X_3$. The minimum probability of transitioning to state $X_3$ under this controller is given likewise by centering stochastic noise at the point $x_{\min}$ furthest from $X_3$ (red point).}
    \label{fig:Transition Probabilities}
    \vspace{-0.2in}
\end{figure}
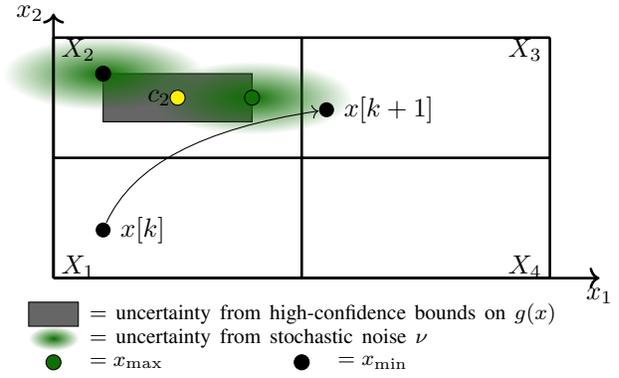

The choice $u[k]=K_{{q'}}(x[k];\hat{g})$ thus produces a control action which compensates for the known and estimated dynamics to reach the center of region $X_{q'}$, although the actual state update will be perturbed as shown in Figure \ref{fig:Transition Probabilities}. 


Our ultimate goal is to apply a sequence of feedback controllers so that the resulting sequence of observations satisfies a control objective specified as a syntactically co-safe LTL (scLTL) formula over the observations $O$.

\begin{definition}[Syntactically co-safe LTL {\cite[Def. 2.3]{belta_formal_2017}}]
A \emph{syntactically co-safe linear temporal logic (scLTL)} formula $\phi$ over a set of observations $O$ is recursively defined as 
\begin{equation*}
    \phi = \true \ |\ o \ |\ \lnot{o} \ |\ \phi_1 \land \phi_2 \ |\ \phi_1 \lor \phi_2 \ |\ \next \phi \ |\ \phi_1\until\phi_2 \ |\ \eventually \phi
\end{equation*}
where $o\in O$ is an observation and $\phi$, $\phi_1$, and $\phi_2$ are scLTL formulas.
We define the \textit{next} operator $\next$ as meaning that $\phi$ will be satisfied in the next state transition, the \textit{until} operator $\until$ as meaning that the system satisfies $\phi_1$ until it satisfies $\phi_2$, 
and the \textit{eventually} operator $\eventually$ as $\true\until\phi$.
\end{definition}
 
ScLTL formulas are characterized by the property that they are satisfied in finite time.
It is well-known that scLTL satisfaction can be checked using a finite state automaton:

\begin{definition}[Finite State Automaton {\cite[Def. 2.4]{belta_formal_2017}}]
A \emph{finite state automaton (FSA)} is a tuple $\mathcal{A} = (S,s_0,O,\delta,F)$, where
\begin{itemize}
    \item $S$ is a finite set of states,
    \item $s_0 \in S$ is the initial state,
    \item $O$ is the input alphabet, which corresponds to observations from the scLTL specification,
    \item $\delta:S\times O\xrightarrow{} S$ is a transition function, and
    \item $F \subseteq S$ is the set of accepting (final) states.
\end{itemize}
\end{definition}
A sequence of inputs (a \textit{word}) from $O$ is said to be accepted by an FSA if it ends in an accepting state. A scLTL formula can always be translated into a FSA that accepts all and only those words satisfying the formula. We use scLTL specifications in this paper because they are well-suited to robotic motion planning tasks which are satisfied in finite time. Additionally, the simpler structure of an FSA as opposed to the B\"{u}chi and Rabin automata of general LTL enables the methods we propose below.
%

\begin{definition}[Interval Markov Decision Process]
An \emph{Interval Markov Decision Process (IMDP)} is a tuple $\mathcal{I} = (Q, A, \check{T}, \hat{T},Q_0, O, L)$ where:
\begin{itemize}
    \item $Q$ is a finite set of states,
    \item $A$ is a finite set of actions,
    \item $\check{T},\hat{T}: Q \times A \times Q' \xrightarrow{} [0,1]$ are lower and upper bounds, respectively, on the transition probability from state $q \in Q$ to state $q' \in Q$ under action $\alpha \in A$,
    \item $Q_0 \subseteq Q$ is a set of initial states,
    \item $O$ is a finite set of atomic propositions or observations,
    \item $L: Q \xrightarrow{} O$ is a labeling function.
\end{itemize}
The set of actions $A$ corresponds to the set of all valid feedback controllers for the system. We do not assume that all actions are available at each state. Therefore, each state has a subset $A(q)\subseteq A$ of available actions.
\end{definition}

\begin{definition}[High-Confidence IMDP Abstraction]
Consider stochastic system \eqref{System Dynamics}, partitions \eqref{eq:part}, and the family of feedback controllers \eqref{eq:feedback} where $\hat{g}(x)$ is an estimate of $g(x)$. Further, suppose that $g_-(x)$ and $g_+(x)$ are high-confidence bounds on $\hat{g}(x)$ which satisfy \eqref{eq:GPbound}. 
Then, an IMDP $\mathcal{I} = (Q, A, \check{T}, \hat{T},Q_0, O, L)$ is a \emph{high-confidence IMDP abstraction of \eqref{System Dynamics}}, 
if:
\begin{itemize}
    \item The set of states $Q$ for the abstraction is the index set of partitions, \textit{i.e.} partition $X_q$ is abstracted as state $q$, and the set of observations $O$ and labeling function $L$ for the abstraction are the same as for the system,
    \item For all $q\in Q$, the set of actions $A(q)$ is the set of one-step reachable regions at $q$ under its feedback controllers,
    \item For all $q\in Q$ and all $\alpha_{q^*}\in A(q)$:
\end{itemize}
\begin{align}
    \label{eq:bound}&\check{T}(q,\alpha_{q^*},q')\leq \\
    \nonumber  &\min_{x\in X_q}\min_{g_-(x)\leq w\leq g_+(x)}\mathbb{P}_\nu(f(x)+w+K_{{q^*}}(x;\hat{g})+\nu\in X_{q'}),\\
    \label{eq:bound2}&\hat{T}(q,\alpha_{q^*},q')\geq \\
    \nonumber &\max_{x\in X_q}\max_{g_-(x)\leq w\leq g_+(x)}\mathbb{P}_\nu(f(x)+w+K_{{q^*}}(x;\hat{g})+\nu\in X_{q'})
\end{align}
\noindent where $\mathbb{P}_\nu$ denotes probability with respect to $\nu$.
\end{definition}


Verification and synthesis problems for IMDP systems evaluated against scLTL specifications are often solved using graph theoretic methods on a product IMDP:
\begin{definition}[PIMDP]\label{def:PIMDP}
Let $\mathcal{I} = (Q, A, \check{T}, \hat{T},Q_0, O, L)$ be an IMDP and $\mathcal{A} = (S,s_0,O,\delta,F)$ be an FSA. The product IMDP (PIMDP) is defined as a tuple $\mathcal{P} = \mathcal{I}\otimes\mathcal{A} =$\\ $(Q\times S, A, \check{T}', \hat{T}', Q\times s_0, F')$, where
\begin{itemize}
    \item $\check{T}': (q,s) \times A \times (q',s') :=
        \check{T}(q,\alpha,q') $ if $ s' \in \delta (s, L(q))$ and
        $0$ otherwise
    \item $\hat{T}': (q,s) \times A \times (q',s') :=
        \hat{T}(q,\alpha,q') $ if $ s' \in \delta (s, L(q))$ and
        $0$ otherwise
    \item $(q_0,\delta(s_0,L(q_0))) \in (Q\times S)$ is a set of initial states of $\mathcal{I}\otimes\mathcal{A}$, and
    \item $F'=Q\times F$ is the set of accepting (final) states.
\end{itemize}
\end{definition}
%

We can now formulate our proposed problem:
\begin{problem}\label{Problem3}
Design an iterative algorithm to sample and learn the unknown dynamics of system \eqref{System Dynamics} without violating the scLTL specification $\phi$ and synthesize a control policy which satisfies $\phi$ with some desired threshold probability or prove that no such control policy exists.
\end{problem}
To solve this problem, we construct a high-confidence IMDP abstraction of the system \eqref{System Dynamics} using a GP estimation of the unknown dynamics. We then formulate a method to sample the state-space without violating the specification, updating the GP estimation until a satisfying control policy can be synthesized.

\section{Abstraction of System as IMDP}\label{section:Abstraction}
In this section, we detail our approach to abstracting a system of the form \eqref{System Dynamics} into a high-confidence IMDP. 

We first need to determine an approximation of $g(x)$, the unknown dynamics. At each time step of system \eqref{System Dynamics}, we know $x[k+1]$, $f(x[k])$, and $u[k]$. Therefore, we can define
\begin{equation*}
    y[k]=x[k+1]-f(x[k])-u[k]=g(x[k])+\nu[k].
\end{equation*}
Then, we construct a Gaussian process estimation $\hat{g}(x)$ for $g(x)$ by considering a dataset of samples $(x[k],y[k])$.

\begin{definition}[Gaussian Process Regression]
\label{def:GP}
Gaussian Process (GP) regression models a function $g_i:\mathbb{R}^n\to \mathbb{R}$ as a distribution with covariance $\kappa:\mathbb{R}^n\times\mathbb{R}^n\xrightarrow{}\mathbb{R}_{>0}$. Assume a dataset of $m$ samples $D = \{(z^j,y_i^j)\}_{j\in\{1,...,m\}}$, where $z^j\in\mathbb{R}^n$ is the input and $y^j_i$ is an observation of $g_i(z^j)$ under Gaussian noise with variance $\sigma_{\nu_i}^2$. 
Let $K\in \mathbb{R}^{m\times m}$ be a matrix defined elementwise by $K_{j\ell}=\kappa(z^j,z^\ell)$ and for $z\in\mathbb{R}^n$, let $k(z)=[\kappa(z,z^1) \;  \kappa(z,z^2) \ldots $ $\kappa(z,z^m)]^T\in \mathbb{R}^m$.
Then, the predictive distribution of $g_i$ at a test point $z$ is the conditional distribution of $g_i$ given $D$, which is Gaussian with mean $\mu_{g_i,D}$ and variance $\sigma_{g_i,D}^2$ given by
\begin{align}\label{mean}
    \mu_{g_i,D}(z) &= k(z)^T(K+\sigma_{\nu_i}^2I_m)^{-1}Y\\
    \label{Std Deviation} \sigma_{g_i,D}^2(z)&=\kappa(z,z)-k(z)^T(K+\sigma_{\nu_i}^2I_m)^{-1}k(z),
\end{align}
where $I_m$ is the identity and $Y=\begin{bmatrix}y^1_i& y^2_i & \ldots & y^m_i\end{bmatrix}^T$.
\end{definition}
In practice, GP regression has a complexity of $O(m^3)$. To mitigate this, we use sparse Gaussian process regression \cite{leibfried2021tutorial}:
\begin{definition}[Sparse Gaussian Process Regression]
A sparse Gaussian process uses a set $D_\eta=\{(z^j,y_i^j)\}_{j\in\{1,...,\eta\}}$ to approximate a GP of a larger dataset $D$. Given \textit{inducing points} $\{z_j\}_{j\in\{1,...,\eta\}}$ with $Y_\eta=\begin{bmatrix}y^1_i& y^2_i & \ldots & y^\eta_i\end{bmatrix}^T$ and covariance matrix $A_\eta$, the predictive distribution of the unknown function $g_i$ has mean $\mu_{g_i,D_\eta}$ and variance $\sigma_{g_i,D_\eta}^2$
\begin{align}\nonumber
    \mu_{g_i,D_\eta}(z) &= k_\eta(z)^T(K_\eta+\sigma_{\nu_i}^2I_\eta)^{-1}Y_\eta\\\nonumber
    \sigma_{g_i,D_\eta}^2(z)&=\kappa(z,z)-k_\eta(z)^TK_\eta^{-1}(K_\eta-A_\eta)K_\eta^{-1}k_\eta(z)
\end{align}
where $K_\eta\in \mathbb{R}^{\eta\times\eta}$ is a matrix defined elementwise by $K_{\eta,j\ell}=\kappa(z^j,z^\ell)$ for all $z\in D_\eta$. For $z\in\mathbb{R}^n$, let $k_\eta(z)=[\kappa(z,z^1) \;  \kappa(z,z^2) \ldots $ $\kappa(z,z^\eta)]^T\in \mathbb{R}^\eta$. The parameters $\{z^j\}_{j\in\{1,...,\eta\}}$, $\{y_i^j\}_{j\in\{1,...,\eta\}}$, and $A_\eta$ are optimized to minimize the Kullback-Leibler divergence (evaluated at the inducing points) between $\mathcal{N}(\mu_{g_i,D_\eta},\sigma_{g_i,D_\eta}^2)$, the posterior of $g_i$ under the sparse GP; and $p(g_{i}|Y)$, the posterior of $g_i$ under a GP with the full dataset $D$. We refer the reader to \cite{leibfried2021tutorial} for a detailed treatment of sparse Gaussian process theory. The computational complexity of sparse GP regression is $O(m\eta^2)$, so by fixing $\eta$ the algorithm is linear in $m$. We note that sparse GP regression introduces error into the estimation; however, in practice this error does not affect the validity of our methods.
\end{definition}
Given some dataset $D$, we construct an estimation of the unknown dynamics independently in each coordinate and determine high-confidence bounds on the estimation error
\begin{align*}
\hat{g}_i^D(x)&:=\mu_{g_i,D}(x),\\
\gamma_i(x)&:=\beta\sigma_{g_i,D}(x)\geq |g_i(x)-\hat{g}_i^{D}(x)|
\end{align*} 
for each $i=1,\ldots,n$. 
We also determine high-confidence lower and upper bounds on $g(x)$ as
\begin{equation*}
    g_-(x)=\hat{g}^D(x)-\beta\sigma_{g,D}(x), \quad g_+(x)=\hat{g}^D(x)+\beta\sigma_{g,D}(x)
\end{equation*}
The parameter $\beta$ is calculated as
\begin{equation}
    \beta=\bigg(\frac{\sigma_\nu}{\sqrt{1+(2/m)}}(B_i+\sigma_\nu\sqrt{2(\gamma_k^m+1+\log\frac{1}{\delta})})\bigg)
\end{equation}
for noise $\sigma_\nu$-sub-Gaussian, $m$ the number of GP samples, high-confidence parameter $\delta$, information gain constant $\gamma_k^m$, and RKHS constant $B_i$ as detailed in Lemma 1, \cite{jackson2021strategy}. Note that the same parameter $\beta\sigma_{g,D}$ is used to determine high-confidence bounds on both the estimation error and $g(x)$ itself.\\
For each region $q$ in the state-space, we select a high-confidence error bound for the unknown dynamics as
\begin{equation*}
    \gamma_i(q)=\max_{x\in X_q}\gamma_i(x)
\end{equation*}
In practice, we compute this bound by sampling $\gamma_i(x)$ throughout the state-space, introducing a trade-off between approximation error and computation complexity.
We now construct transition probability intervals assuming that the high-confidence bounds on unknown dynamics always hold:

\begin{thm}[Construction of Transition Probabilities] \label{thm:ProbInterval}

Consider $q,q'\in Q$ and action $\alpha_{q^*}\in A(q)$. Lower bound $\check{T}$ and upper bound $\hat{T}$ transition probabilities from $q$ to $q'$ under $\alpha_{q^*}$ are given by
\begin{equation}\label{eq:lowerTBound}
    \check{T}(q,\alpha_{q^*},q')=\prod_{i=1}^n\int_{a'_i}^{b'_i}\rho_{\nu_i}(z-x_{\min,i}(q,\alpha_{q^*},q'))dz,
\end{equation}
\begin{equation}\label{eq:upperTBound}
    \hat{T}(q,\alpha_{q^*},q')=\prod_{i=1}^n\int_{a'_i}^{b'_i}\rho_{\nu_i}(z-x_{\max,i}(q,\alpha_{q^*},q'))dz,
\end{equation}
where $x_{\min,i}$ is the $i$-th coordinate of $x_{\min}$  and similarly for $x_{\max,i}$, we recall $\rho_{\nu_i}$ is the probability density function of the stochastic noise $\nu_i$, and $a'$ and $b'$ are the lower and upper boundary points for region $q'$. We define $x_{\min}$ and $x_{\max}$ as
\begin{align}
    \label{eq:xmin} x_{\min}(q,\alpha_{q^*},q')= \ &\underset{x\in X}{\mathrm{argmax}}\ ||x-c_{q'}||_1 \\\nonumber
    &\text{s.t.}\ c_{q^*}-\gamma(q)\leq x\leq c_{q^*}+\gamma(q),
\end{align}
\begin{align}
 \label{eq:xmax}    x_{\max}(q,\alpha_{q^*},q')= \ &\underset{x\in X}{\mathrm{argmin}}\ ||x-c_{q'}||_1 \\\nonumber
    &\text{s.t.}\ c_{q^*}-\gamma(q)\leq x\leq c_{q^*}+\gamma(q),
\end{align}
where $||\cdot||_1$ is the 1-norm and $\gamma(q)$ is a high-confidence error bound on the unknown dynamics satisfying Assumption \ref{assum:1}. \\
Then, the transition probability bounds
\eqref{eq:lowerTBound}--\eqref{eq:upperTBound} satisfy the constraints for high-confidence IMDP abstractions in \eqref{eq:bound}--\eqref{eq:bound2}. \\
\textit{Proof:} \quad
The righthand side of the bound in equation \eqref{eq:bound} can be rewritten as
\begin{align}
    \label{eq:chain1} &\min_{x\in X_q}\min_{\substack{g_-(x)\leq w\\\leq g_+(x)}}\mathbb{P}_\nu(f(x)+w+K_{{q^*}}(x;\hat{g})+\nu\in X_{q'})\\
    \label{eq:chain2} &=\min_{x\in X_q}\min_{\substack{g_-(x)\leq w\leq g_+(x)}}\mathbb{P}_\nu(c_{{q^*}}+w-\hat{g}(x)+\nu\in q')\\
    \label{eq:chain3} &= \min_{x\in X_q}\min_{\substack{-\gamma(x)\leq \omega\leq \gamma(x)}}\mathbb{P}_\nu(c_{{q^*}}+\omega+\nu\in X_{q'})\\
\label{eq:chain4} &=\min_{x\in X_q}\min_{\substack{-\gamma(x)\leq \omega\\\leq \gamma(x)}}\prod_{i=1}^n\mathbb{P}_{\nu_i}(c_{{q^*},i}+\omega_i+\nu_i\in [a_{q',i},b_{q',i}])\\
\label{eq:chain5} &=\min_{x\in X_q}\prod_{i=1}^n\min_{\substack{-\gamma_i(x)\leq \omega_i\\\leq \gamma_i(x)}}\mathbb{P}_{\nu_i}(c_{{q^*},i}+\omega_i+\nu_i\in [a_{q',i},b_{q',i}])\\
\label{eq:chain6} &\geq \prod_{i=1}^n\min_{\substack{-\gamma_i(q)\leq \omega_i\\\leq \gamma_i(q)}}\mathbb{P}_{\nu_i}(c_{{q^*},i}+\omega_i+\nu_i\in [a_{q',i},b_{q',i}])
\end{align}
where \eqref{eq:chain1} is the righthand side of \eqref{eq:bound}; \eqref{eq:chain2} follows after expanding the feedback controller expression $K_{{q^*}}(x;\hat{g})$ using \eqref{eq:feedback} and simplifying; \eqref{eq:chain3} follows 
by assumption of high-confidence error bound $\gamma(x)$ and the definition of $g_-(x)$ and $g_+(x)$ from Assumption \ref{assum:1} and taking $\omega=w-\hat{g}(x)$; \eqref{eq:chain4} follows by assumption that each $\nu_i$ is independent and $\mathbb{P}_{\nu_i}$ denotes probability with respect to $\nu_i$, where we recall that $a_{q'}$ and $b_{q'}$ are the lower and upper corners of the region $X_{q'}$, and $a_{q',i}$ is the $i$-th coordinate of $a_{q'}$ and similarly for $c_{q*,i}$ and $b_{q',i}$;  \eqref{eq:chain5} follows from the fact that the hyper-rectangular constraint $-\gamma(x)\leq \omega\leq \gamma(x)$ is equivalent to independent constraint $-\gamma_i(x)\leq \omega_i\leq \gamma_i(x)$ along each coordinate; and \eqref{eq:chain6} follows from the definition $\gamma_i(q)=\max_{x\in X_q}\gamma_i(x)$. 

Now, because the probability distribution for each random variable $\nu_i$ is assumed unimodal and symmetric, $\mathbb{P}_{\nu_i}(c_{{q^*},i}+\omega_i+\nu_i\in [a_{q',i},b_{q',i}])$ is minimized when the distance between $(c_{{q^*},i}+\omega_i)$ and the midpoint of $[a_{q',i},b_{q',i}]$ is maximized, \emph{i.e.}, when $|c_{{q^*},i}+\omega_i-c_{q',i}|$ is maximized, subject to the constraint $-\gamma_{i}(q)\leq\omega_i\leq\gamma_i(q)$. Substituting $x=c_{q^*}+\omega$, and observing that $\|x-c_{q'}\|_1=\sum_{i=1}^n|x_i-c_{q',i}|$, this is exactly the maximizing point specified by $x_{\min}(q,\alpha_{q^*},q')$ in \eqref{eq:xmin}.  Thus, the expression in \eqref{eq:chain6} is equivalent to
\begin{align}
    \prod_{i=1}^n\mathbb{P}_{\nu_i}(x_{\min,i}(q,\alpha_{q^*},q')+\nu_i\in [a_{q',i},b_{q',i}]),
\end{align}
which in turn is equivalent to the righthand side of \eqref{eq:lowerTBound}, establishing the bound in \eqref{eq:bound}. An analogous argument establishes that \eqref{eq:upperTBound} satisfies \eqref{eq:bound2}.
\QED
\end{thm}


We construct a high-confidence IMDP abstraction of the system using the hyper-rectangular partition regions as states, high-confidence bounds on the unknown dynamics obtained via GP regression, and transition probability intervals calculated using Theorem \ref{thm:ProbInterval}, solving the first part of Problem \ref{Problem3}.

\section{Safe Sampling of PIMDP}\label{section:Sampling}

\subsection{Probability of Satisfaction Calculation}
Given a high-confidence IMDP abstraction of the system and a FSA of a desired scLTL specification, we construct a PIMDP using Definition \ref{def:PIMDP}. We first introduce the concept of control policies and adversaries:
\begin{definition}[Control Policy]
A control policy $\pi\in\Pi$ of a PIMDP is a mapping $(Q\times S)^+\xrightarrow{}A$, where $(Q\times S)^+$ is the set of finite sequences of states of the PIMDP.
\end{definition}
\begin{definition}[PIMDP Adversary]
Given a PIMDP state $(q,s)$ and action $\alpha$, an adversary $\xi\in\Xi$ is an assignment of transition probabilities $T_{\xi}'$ to all states $(q',s')$ such that
\begin{align*}
    \check{T}'((q,s),\alpha,(q',s'))&\leq T_{\xi}'((q,s),\alpha,(q',s')) \\ &\leq\hat{T}'((q,s),\alpha,(q',s')).
\end{align*}
In particular, we use a \textit{minimizing} adversary, which realizes transition probabilities such that the probability of satisfying the specification is minimal, and a \textit{maximizing} adversary, which maximizes the probability of satisfaction.
\end{definition}
To find safe sampling cycles in the PIMDP, we calculate
\begin{equation*}
    \check{P}_{\max}((q,s)\hspace{-2pt}\models\hspace{-2pt}\phi)=\max\limits_{\pi\in\Pi}\min\limits_{\xi\in\Xi}P(w\hspace{-2pt}\models\hspace{-2pt}\phi \ |\ \pi,\xi,w[0]=(q,s)),
\end{equation*}
which is the probability that a random path $w$ starting at PIMDP state $(q,s)$ satisfies the scLTL specification $\phi$ under a maximizing control policy $\pi$ and minimizing adversary $\xi$.

Additionally, we will also use the best case probability of satisfaction under a maximizing control policy and adversary:
\begin{equation*}
    \hat{P}_{\max}((q,s)\hspace{-2pt}\models\hspace{-2pt}\phi)=\max\limits_{\pi\in\Pi}\max\limits_{\xi\in\Xi}P(w_i\hspace{-2pt}\models\hspace{-2pt}\phi \ |\ \pi,\xi,w[0]=(q,s))
\end{equation*}
To calculate these probabilities, we use a value iteration method proposed in Section V, \cite{lahijanian_formal_2015}.

\subsection{Nonviolating Sub-Graph Generation}

We note that scLTL specifications may generate FSA states which are absorbing and non-accepting, i.e., it is impossible to satisfy the specification once one of these states is reached. Such states may also exist in PIMDP constructions even without appearing in the corresponding FSA. We define these states as those which have zero probability of satisfying the scLTL specification under any control policy and adversary:
\begin{equation*}
    \text{Failure States}
    =\{(q,s)\in Q\times S\ |\ \hat{P}_{\max}((q,s)\models\phi)=0\}.
\end{equation*}

We can then define a notion of specification nonviolation:
\begin{definition}[Nonviolating PIMDP]\label{def:PIMDP Safety}
A PIMDP $\mathcal{P}$ is \textit{nonviolating} with respect to a scLTL specification $\phi$ if there exists no failure states in $\mathcal{P}$.
\end{definition}

Our algorithm for calculating a nonviolating PIMDP is as follows. 
We first initialize a set of failure states. Then, 
we loop through all non-failure states and prune actions which have nonzero upper-bound transition probability to failure states. 
We check if this pruning has left any states with no available actions, designating these also as failure states to prune. The process continues until no new failure states are found. Our nonviolating sub-graph is the set of all unpruned states with their remaining actions.

\subsection{Candidate Cycle Selection}
Now that we have a nonviolating sub-graph of our PIMDP, we want to select a path which we can take in order to sample the state-space indefinitely while maximizing the information gain of our Gaussian process. To do this, we first recall the concept of \textit{maximal end components} \cite{baier_principles_2008}:

\begin{definition}[End Component \cite{baier_principles_2008}]
An \textit{end component} of a finite PIMDP $\mathcal{P}$ is a pair $(\mathcal{T},Act)$ with $\mathcal{T}\subseteq(Q\times S)$ and $Act:\mathcal{T}\rightarrow A$ such that
\begin{itemize}
    \item $\emptyset\neq Act(q,s)\subseteq A(q)$ for all states $(q,s)\in\mathcal{T}$,
    \item $(q,s)\in\mathcal{T}$ and $\alpha\in Act(q,s)$ implies $\{(q',s')\in\mathcal{T}\ |\ \hat{T}(q,\alpha,q'))>0,s'\in\delta(s,L(q))\}\subseteq\mathcal{T}$,
    \item The digraph $G_{(\mathcal{T},Act)}$ induced by $(\mathcal{T},Act)$ is strongly connected.
\end{itemize}
\end{definition}
\begin{definition}[Maximal End Component (MEC) \cite{baier_principles_2008}]
An end component $(\mathcal{T},Act)$ of a finite PIMDP $\mathcal{P}$ is \textit{maximal} if there is no end component $(\mathcal{T}^*,Act^*)$ such that $(\mathcal{T},Act)\neq (\mathcal{T}^*,Act^*)$ and $\mathcal{T}\subseteq \mathcal{T}^*$ and $Act(q,s)\subseteq Act^*(q,s)$ for all $(q,s)\in \mathcal{T}$.
\end{definition}

PIMDP abstractions have the property that any infinite path will eventually stay in a single MEC.
We propose the following heuristic in order to select a MEC to cycle within. First, we calculate $\check{P}_{\max}$ from our initial state to each candidate MEC. 
We reject any MEC which we cannot reach with probability 1, or, in case no MECs can be reached with probability 1, we immediately select the MEC with the highest reachability probability. If multiple candidate MECs remain, we then calculate the Gaussian process covariance $\kappa(c_q,c_{q^*})$ between the centers of the IMDP states $q$ in each remaining candidate MEC and the accepting IMDP state $q^*$. We sum the covariances for all states in each MEC and select the MEC with the highest total covariance score, which corresponds to maximum information gain \cite{srinivas_regret}, defined as reduction of GP uncertainty at the accepting state. We generate a control policy by selecting the actions at each state which give the maximum probability of reaching the MEC. Once in the MEC, we use a controller which cycles through the available actions.

By applying the algorithms detailed above to calculate a non-violating PIMDP and MEC, we generate a control policy which samples the state-space indefinitely without violating the specification, solving the second part of Problem \ref{Problem3}.

\subsection{Iterative Sampling Algorithm}
We now detail our complete method to solve Problem \ref{Problem3}. Given a scLTL specification $\phi$ which we want to satisfy with probability $P_{\text{sat}}$, we construct a PIMDP using a high-confidence IMDP abstraction of the system in Eq.~(\ref{System Dynamics}) and an FSA which models $\phi$. Then, we calculate reachability probabilities under a minimizing adversary $\check{P}_{\max}$ from the initial states in the PIMDP to the accepting states. If $\check{P}_{\max}\geq P_{\text{sat}}$, then the control policy selects the actions which produce $\check{P}_{\max}$ at each state and the problem is solved. Otherwise, we 
calculate a control policy to sample the state-space without violating the specification $\phi$ using the methods in previous sections. We follow the calculated control policy for a predetermined number of steps and sample the unknown dynamics at each step. We batch update the GP with the data collected, reconstruct transition probability intervals for each state, and recalculate reachability probabilities $\check{P}_{\max}$ for our initial states. If $\check{P}_{\max}\geq P_{\text{sat}}$, a satisfying control policy is found; otherwise, we repeat the process above. Our iterative algorithm ends when $\check{P}_{\max}\geq P_{\text{sat}}$; the GP approximation has low enough uncertainty to know that a successful control policy cannot be synthesized, \textit{i.e.}, when the reachability probability $\hat{P}_{\max}$ under a maximizing adversary is less than the desired $P_{\text{sat}}$; or a maximum number of iterations has been reached.

\section{Case Study}
Suppose we have a mobile robot in a 2D state-space with position $x\in X:=[0,5]^2\subset \mathbb{R}^2$. The state-space is partitioned into a set of 25 hyper-rectangular regions corresponding to IMDP states. The dynamics of the robot are
\begin{equation}
    x[k+1] = x[k] + u[k] + g(x[k]) + v
\end{equation}
where $g(x)$ models the unknown effect of the slope of the terrain. 
The control action $u$ is generated by the family of controllers in Section \ref{section:Setup} where the set of available target regions are those left, right, above, or below each region.

Within the state-space, we have one goal region with the atomic proposition \texttt{Goal} and a set of hazard regions labeled with \texttt{Haz}. These yield the scLTL specification
\begin{equation}
    \phi_1 = \lnot \texttt{Haz} \ \until \ \texttt{Goal}.
\end{equation}

\begin{figure}[t]
    \centering
    \includegraphics[width=0.225\textwidth]{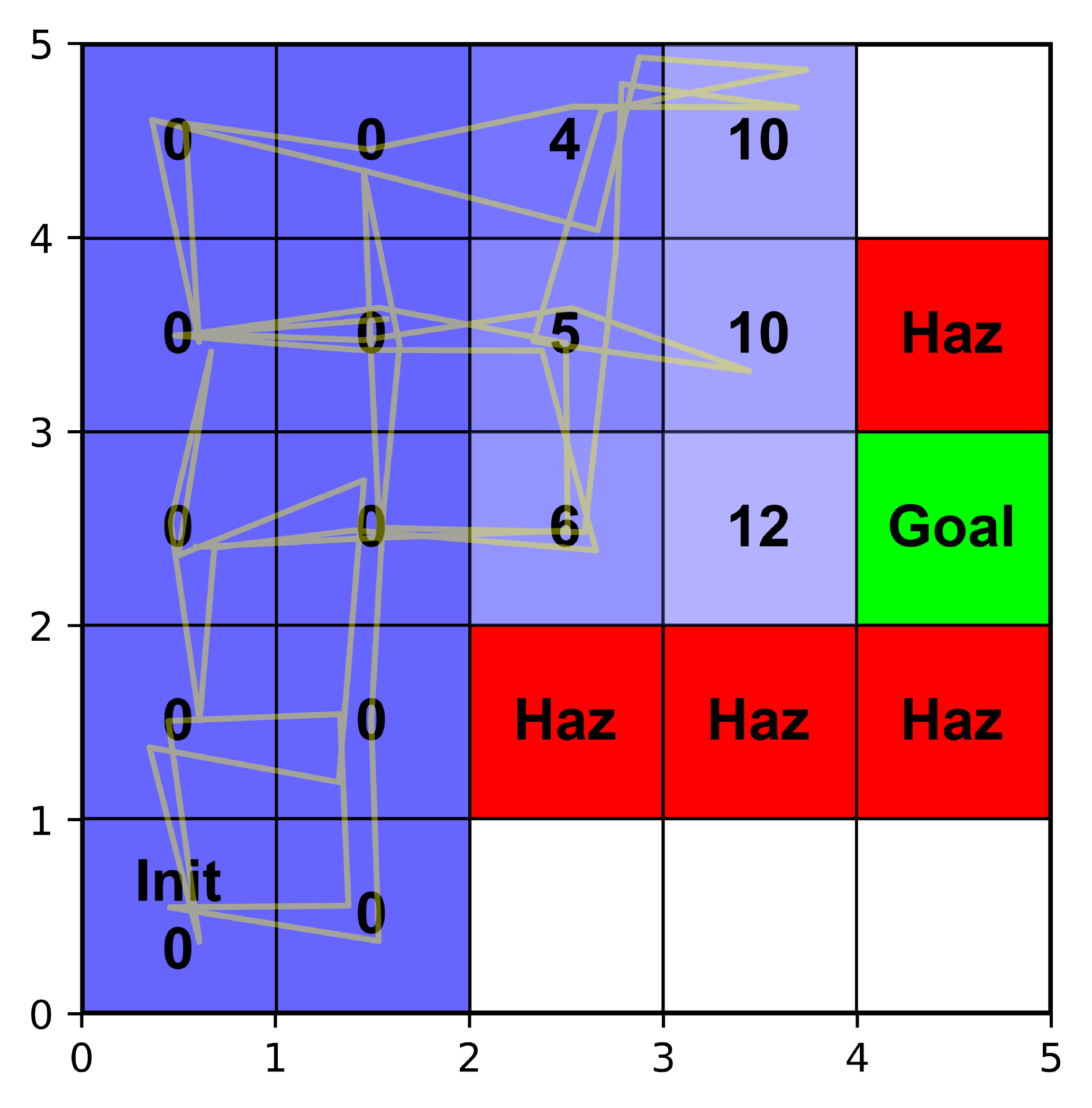}
    \caption{State-space of the case study. The initial region is labeled with "Init", the (green) target region is labeled with "Goal", and the (red) hazard regions are labeled with "Haz". States that eventually enter the safe cycle are blue, and the number in the region indicates the iteration of the algorithm at which the state enters the safe cycle. States which are not numbered do not enter the safe cycle. The yellow trace is an example of a sampling run.}
    \label{fig:Safe Cycle}
    \vspace{-0.2in}
\end{figure}

An illustration of the state-space is shown in Figure \ref{fig:Safe Cycle}. We choose a low-dimensional case study in order to illustrate our methodology. Future works will refine our algorithms on applications with higher-dimensional state-spaces.

The true $g(x)$ is sampled from two randomly generated Gaussian processes (one for each dimension) with bounded support $[-0.4,0.4]$ and {squared exponential} kernel $\kappa$,
\begin{equation}
    \kappa(x,x')=\sigma_g^2e^{-\frac{(x-x')^2}{2l^2}}.
\end{equation}
We choose hyperparameters $\sigma_g=0.45$ and $l=1.75$.

We estimate the unknown dynamics with two sparse Gaussian processes with the same kernel as the true dynamics. We sample the GPs at 100 points in each region to determine error bounds. 
We set the number of inducing points $\eta=250$ and choose our high-confidence-bound parameter $\beta=2$. Each iteration of the algorithm takes 250 steps, so the total number of data samples $m$ is the number of iterations times 250.
Our stochastic noise $\nu$ is independently drawn from two truncated Gaussian distributions, one for each dimension, and both with $\sigma_\nu = 0.1$ and bounded support $[-0.2,0.2]$.


We next apply the iterative algorithm described in Section IV-D, setting the desired probability of satisfying the specification to 1. Our algorithm successfully finds a satisfying feedback control strategy in an average of 15 iterations (calculated over 10 runs). The algorithm is implemented in Python on a 2.5 GHz Intel Core i9 machine with 16 GB of RAM and a Nvidia RTX 3060 GPU, and requires on average 1 minute 14 seconds to complete.

Figure \ref{fig:Safe Cycle} depicts the expansion of the safe cycle used to sample the state-space. Initially, only the left two columns of states are safe and reachable. As the algorithm progresses, more states and actions are added to the safe cycle, moving the system closer to the goal until the unknown dynamics can be estimated with enough certainty to achieve a probability of satisfying the specification of 1.

\begin{figure}[t]
    \centering
    \includegraphics[width=0.45\textwidth]{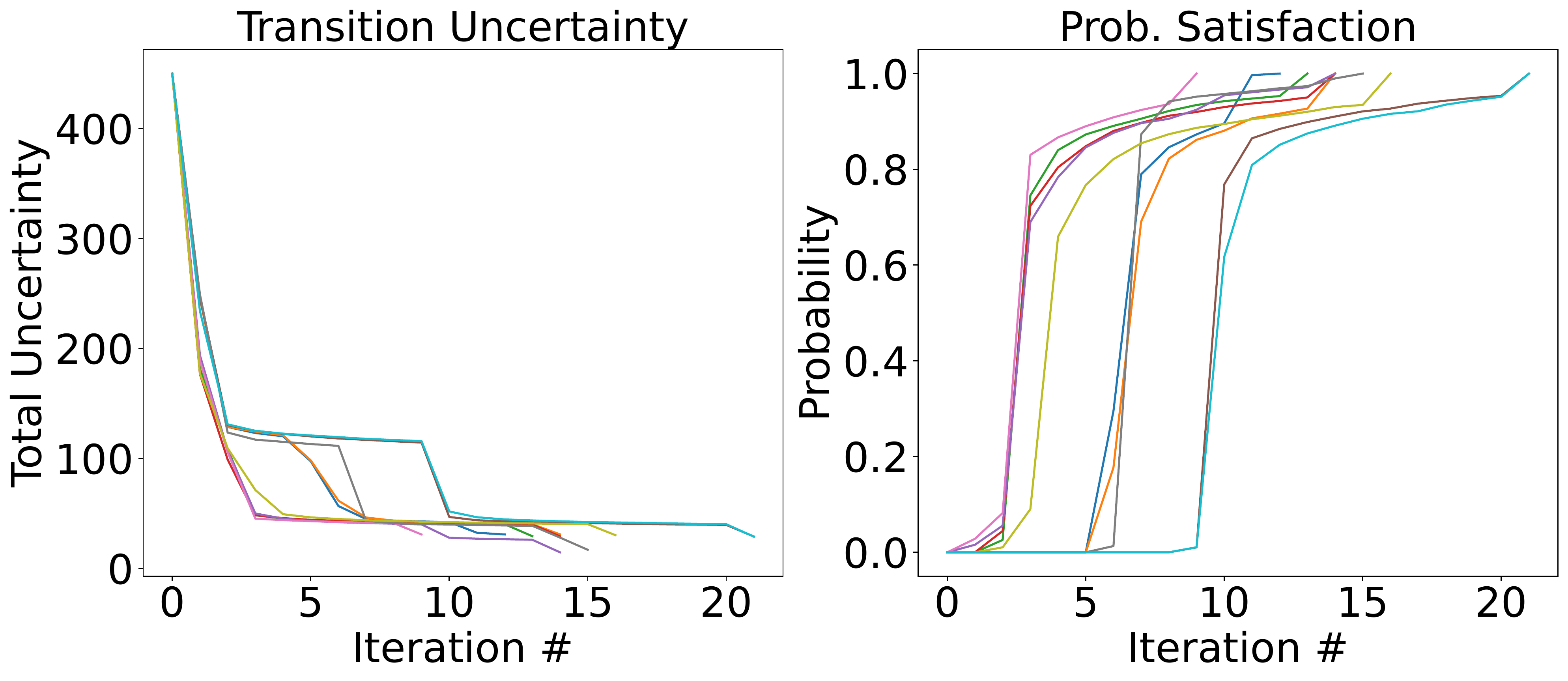}
    \caption{The left plot shows the total uncertainty in transition probability intervals after each iteration of the algorithm, and the right plot shows the probability of satisfying the specification after each iteration. Results are plotted over 10 runs of the algorithm. The uncertainty decreases as more data samples are collected, and likewise the probability of satisfaction increases once the safe cycle has expanded close enough to the goal.}
    \label{fig:Metrics}
    \vspace{-0.2in}
\end{figure}

The left plot in Figure \ref{fig:Metrics} depicts the total transition probability uncertainty for the system after each iteration
\begin{equation}\label{eq:TUncertainty}
    T_{{\rm unc}, {\rm total}}=\sum_{q\in Q}\sum_{\alpha\in A(q)}\sum_{q'\in Q}\hat{T}(q,\alpha,q')-\check{T}(q,\alpha,q').
\end{equation}

The right plot in Figure \ref{fig:Metrics} shows the probability of satisfying the specification after each iteration.

\section{Conclusion}
In this work, we developed a method to safely learn unknown dynamics for a system motivated by the robotic motion-planning problem. Our approach uses an IMDP abstraction of the system and a finite state automaton of scLTL specifications. We designed an algorithm for finding nonviolating paths within a product IMDP construction which can be used to sample the state-space and construct a Gaussian process approximation of unknown dynamics. We then detailed an algorithm to iteratively sample the state-space to improve the probability of satisfying a desired specification and demonstrated its use with a case study of robot navigation. Our approach can be used with any system for which a high-confidence IMDP abstraction can be constructed as well as any objective which can be written as a scLTL specification. Future work will apply these methods to models of bipedal walking robots utilizing region-based motion planning \cite{shamsah_integrated_2021}. 


\bibliographystyle{IEEEtran}
\bibliography{references, ref}

\begin{thebibliography}{10}
\providecommand{\url}[1]{#1}
\csname url@samestyle\endcsname
\providecommand{\newblock}{\relax}
\providecommand{\bibinfo}[2]{#2}
\providecommand{\BIBentrySTDinterwordspacing}{\spaceskip=0pt\relax}
\providecommand{\BIBentryALTinterwordstretchfactor}{4}
\providecommand{\BIBentryALTinterwordspacing}{\spaceskip=\fontdimen2\font plus
\BIBentryALTinterwordstretchfactor\fontdimen3\font minus
  \fontdimen4\font\relax}
\providecommand{\BIBforeignlanguage}[2]{{%
\expandafter\ifx\csname l@#1\endcsname\relax
\typeout{** WARNING: IEEEtran.bst: No hyphenation pattern has been}%
\typeout{** loaded for the language `#1'. Using the pattern for}%
\typeout{** the default language instead.}%
\else
\language=\csname l@#1\endcsname
\fi
#2}}
\providecommand{\BIBdecl}{\relax}
\BIBdecl

\bibitem{Tabuada2009}
P.~Tabuada, \emph{Verification and Control of Hybrid Systems: A Symbolic
  Approach}, 1st~ed.\hskip 1em plus 0.5em minus 0.4em\relax Springer Publishing
  Company, Incorporated, 2009.

\bibitem{belta_formal_2017}
C.~Belta, B.~Yordanov, and E.~Göl, \emph{\BIBforeignlanguage{en}{Formal
  {Methods} for {Discrete}-{Time} {Dynamical} {Systems}}}, ser. Studies in
  {Systems}, {Decision} and {Control}.\hskip 1em plus 0.5em minus 0.4em\relax
  Springer International Publishing, 2017.

\bibitem{Mark_2020}
C.~Mark and S.~Liu, ``Stochastic {MPC} with distributionally robust chance
  constraints,'' \emph{IFAC}, vol.~53, no.~2, pp. 7136--7141, 2020.

\bibitem{Lavaei2021}
A.~Lavaei, S.~Soudjani, A.~Abate, and M.~Zamani, ``Automated verification and
  synthesis of stochastic hybrid systems: A survey,'' 2021.

\bibitem{baier_model-checking_2003}
C.~Baier, B.~Haverkort, H.~Hermanns, and J.-P. Katoen, ``Model-checking
  algorithms for continuous-time {Markov} chains,'' \emph{IEEE Transactions on
  Software Engineering}, vol.~29, no.~6, pp. 524--541, Jun. 2003.

\bibitem{ahmadi-safety-2017}
M.~Ahmadi, A.~Israel, and U.~Topcu, ``Safety assessemt based on
  physically-viable data-driven models,'' in \emph{56th IEEE {CDC}}, Dec. 2017,
  pp. 6409--6414.

\bibitem{jackson2021strategy}
J.~Jackson, L.~Laurenti, E.~Frew, and M.~Lahijanian, ``Strategy synthesis for
  partially-known switched stochastic systems,'' in \emph{Proceedings of HSCC
  '21}, pp. 1--11.

\bibitem{Williams96gaussianprocesses}
C.~K.~I. Williams and C.~E. Rasmussen, ``Gaussian processes for regression,''
  in \emph{Advances in neural information processing systems 8}.\hskip 1em plus
  0.5em minus 0.4em\relax MIT press, 1996, pp. 514--520.

\bibitem{akametalu_reachability-based_2014}
A.~K. Akametalu, J.~F. Fisac, J.~H. Gillula, S.~Kaynama, M.~N. Zeilinger, and
  C.~J. Tomlin, ``Reachability-based safe learning with {Gaussian} processes,''
  in \emph{53rd {IEEE} {CDC}}, Dec. 2014, pp. 1424--1431.

\bibitem{leibfried2021tutorial}
F.~Leibfried, V.~Dutordoir, S.~John, and N.~Durrande, ``A tutorial on sparse
  gaussian processes and variational inference,'' 2021, arXiv: 2012.13962.

\bibitem{Julius2008}
A.~A. Julius, A.~Halasz, M.~S. Sakar, H.~Rubin, V.~Kumar, and G.~J. Pappas,
  ``Stochastic modeling and control of biological systems: The lactose
  regulation system of \textit{Escherichia Coli},'' \emph{IEEE Transactions on
  Automatic Control}, vol.~53, no. Special Issue, pp. 51--65, 2008.

\bibitem{ALTMAN19991937}
E.~Altman, T.~Başar, and R.~Srikant, ``Congestion control as a stochastic
  control problem with action delays,'' \emph{Automatica}, vol.~35, no.~12, pp.
  1937--1950, 1999.

\bibitem{shamsah_integrated_2021}
A.~Shamsah, J.~Warnke, Z.~Gu, and Y.~Zhao, ``Integrated {Task} and {Motion}
  {Planning} for {Safe} {Legged} {Navigation} in {Partially} {Observable}
  {Environments},'' 2021, arXiv: 2110.12097.

\bibitem{lahijanian_formal_2015}
M.~Lahijanian, S.~B. Andersson, and C.~Belta, ``Formal {Verification} and
  {Synthesis} for {Discrete}-{Time} {Stochastic} {Systems},'' \emph{IEEE
  Transactions on Automatic Control}, vol.~60, no.~8, pp. 2031--2045, Aug.
  2015.

\bibitem{baier_principles_2008}
C.~Baier and J.-P. Katoen, \emph{\BIBforeignlanguage{en}{Principles of {Model}
  {Checking}}}.\hskip 1em plus 0.5em minus 0.4em\relax MIT Press, 2008.

\bibitem{srinivas_regret}
N.~Srinivas, A.~Krause, S.~M. Kakade, and M.~W. Seeger, ``Information-theoretic
  regret bounds for gaussian process optimization in the bandit setting,''
  \emph{IEEE Transactions on Information Theory}, vol.~58, no.~5, p.
  3250–3265, May 2012.

\end{thebibliography}

\appendix[Algorithm Psuedocodes]
Algorithm \ref{Sub-Graph Generation} calculates a sub-graph of a PIMDP which is nonviolating with respect to a scLTL specification. It takes as input a PIMDP construction along with upper bounds $\hat{P}_{\rm max}$ on the maximum probability of satisfying the specification for each state. In lines 1--2, failure states which have $\hat{P}_{\rm max}=0$ are identified. Next, in lines 4--10, non-failure states are looped through and any of their actions which have a nonzero upper bound probability of reaching the set of failure states are removed. In lines 12--17, states which have no actions remaining are added to the set of failure states and the algorithm returns to line 3. The loop from lines 3--18 repeats until no new failure states are identified. The algorithm returns the remaining non-failure states and actions as the nonviolating sub-graph.
\begin{algorithm}
\caption{Nonviolating Sub-Graph Generation}\label{Sub-Graph Generation}
\KwIn{PIMDP $\mathcal{P}$, $\hat{P}_{\max}$ for each state in $\mathcal{P}$}
\KwOut{PIMDP $\mathcal{P}$' which is a nonviolating subset of $\mathcal{P}$}
\SetKw{Init}{Initialize}
\Init{$R = \{(q,s)\in \mathcal{P}\ |\ \hat{P}_{\max}(q,s\models\phi)=0\}$}\;
\Init{$U = R$}\;
\While{$R \neq \emptyset$}{
    \For{$(q,s)\in \mathcal{P}\setminus U$}{
        \For{$\alpha\in A$}{
            \If{$\hat{T}((q,s),\alpha,U)\neq 0$}{
                Remove $\alpha$ from available actions at $(q,s)$ \;
            }
        }
    }
    $R = \emptyset$ \;
    \For{$(q,s)\in \mathcal{P}\setminus U$}{
        \If{$A((q,s))=\emptyset$}{
            $R = R \cup (q,s)$ \;
            $U = U \cup (q,s)$ \;
        }
    }
}
\KwRet{$\mathcal{P}' = \mathcal{P}\setminus U$}
\end{algorithm}
\\
Algorithm \ref{Safe Cycle Selection} calculates a maximum end component of a PIMDP to sample along with a corresponding control policy, maximizing the information gain with respect to learning the unknown dynamics. It takes as input a nonviolating sub-graph of a PIMDP. In line 1, the maximal end components of the sub-graph are identified. In lines 3--8, a lower bound $\check{P}_{\rm max}$ on the maximum probability of reaching each MEC from the initial state of the original PIMDP is calculated. Those MECs which have $\check{P}_{\max}=1$ are added to the list of candidate MECs. In lines 9--11, if there are no candidate MECs found, then the algorithm selects the MEC with the highest $\check{P}_{\rm max}$ as the MEC to cycle in. If there are candidate MECs, then in lines 12--17 each candidate MEC is assigned a score equal to the sum of the covariances between each state in the MEC and the accepting state $q^*$ of the PIMDP. The MEC with the maximum covariance score is selected as the MEC to cycle in. In line 18, a control policy for the selected MEC is calculated which selects the actions at each state outside the MEC which have maximum probability of reaching the MEC. For states within the MEC, the control policy cycles through the actions at each state which are available in the MEC. The algorithm returns the selected MEC along with its corresponding control policy.
\begin{algorithm}
\caption{Nonviolating Cycle Selection}\label{Safe Cycle Selection}
\KwIn{Nonviolating Sub-PIMDP $\mathcal{P}'$}
\KwOut{Selected MEC $(\mathcal{T}^*,Act^*)$ and control policy $\pi^*$}
\SetKw{Init}{Initialize}
\Init{$M$ as the MECs of $\mathcal{P}'$}\;
\Init{$C=\emptyset$ as the set of candidate MECs}\;
\For{$(\mathcal{T}^\dagger,Act^\dagger) \in M$}{
    Calculate reachability probability $\check{\mathcal{P}}_{\max}$ from initial state $r_0$ of PIMDP $\mathcal{P}'$ to $\mathcal{T}^\dagger$\;
    \If{$\check{P}_{\max}=1$}{
        $C=C\cup(\mathcal{T}^\dagger,Act^\dagger)$\;
    }
}
\If{$C=\emptyset$}{
    $(\mathcal{T}^*,Act^*)=\underset{(\mathcal{T}^\dagger,Act^\dagger) \in M}{\mathrm{argmax}}\check{P}_{\max}(r_0\models\eventually \mathcal{T}^\dagger)$\;
} \Else{
    \For{$(\mathcal{T}^\dagger,Act^\dagger) \in C$}{
        $H$ = Sum $\kappa(c_q,c_{q^*})$ for all IMDP states $q\in \mathcal{T}^\dagger$ w.r.t. accepting state $q^*$\;
    }
    Find $(\mathcal{T}^*,Act^*) \in C$ with maximum score $H$\;
}
Control policy $\pi^*$ selects available actions in $\mathcal{P}'$ with maximum probability of reaching $\mathcal{T}^*$ for states not in $\mathcal{T}^*$ and cycles through actions $Act^*$ for all states in $\mathcal{T}^*$\;
\KwRet{Selected MEC $(\mathcal{T}^*,Act^*)$, control policy $\pi^*$}
\end{algorithm}
\\
Algorithm \ref{Iterative Synthesis Algorithm} performs an iterative procedure to safely learn the unknown dynamics of the system \eqref{System Dynamics} until a given scLTL specification can be satisfied with sufficient probability. It takes as input the system dynamics, a scLTL specification, and a desired probability of satisfaction $P_{\rm sat}$. In line 1, an IMDP abstraction of the system and a FSA of the specification are constructed. Then, the IMDP and FSA are combined into a PIMDP construction. Finally, a GP estimation $\hat{g}(x)$ of the unknown dynamics is initialized with its hyperparameters. In lines 2--3, lower and upper bounds $\check{P}_{\rm max}$ and $\hat{P}_{\rm max}$ are calculated for the initial state in the PIMDP. If the lower bound probability $\check{P}_{\rm max}$ is less than the desired $P_{\rm sat}$, the loop in lines 4--18 is entered. In lines 5--7, if the upper bound probability $\hat{P}_{\rm max}$ is less than $P_{\rm sat}$, then the specification cannot be satisfied with sufficient probability regardless of how well the unknown dynamics are learned. Thus, the algorithm returns that no satisfying control policy exists. Otherwise, in lines 8--9, a nonviolating sub-graph of the PIMDP is calculated using Algorithm \ref{Sub-Graph Generation} and a MEC to cycle in along with its corresponding control policy is calculated from this sub-graph using Algorithm 2. In lines 10--13, this control policy is used to take a predefined number of steps to sample the unknown dynamics. In lines 14--15, the GP estimation $\hat{g}(x)$ is updated using these samples, and transition probability intervals are recalculated for each state in the PIMDP. In lines 16--17, $\check{P}_{\rm max}$ and $\hat{P}_{\rm max}$ are recalculated for the initial state. If $\check{P}_{\rm max}\geq P_{\rm sat}$, the loop terminates and the algorithm returns a control policy calculated in lines 19--21 which selects the actions at each state which have maximum probability of satisfying the specification. If $\check{P}_{\rm max}<P_{\rm sat}$, the loop repeats from line 4. If the maximum number of iterations of the loop is reached, the algorithm terminates without determining a satisfying control policy or the nonexistence thereof.
\begin{algorithm}
\caption{Iterative Synthesis Algorithm}\label{Iterative Synthesis Algorithm}
\KwIn{System dynamics in~(\ref{System Dynamics}), scLTL specification $\phi, P_{\text{sat}}$}
\KwOut{Satisfying control policy $\pi^\dagger$ or proof of nonexistence}
\SetKw{Construct}{Construct}
\Construct IMDP $\mathcal{I}$ from System \ref{System Dynamics},
FSA $\mathcal{A}$ from $\phi$,
PIMDP $\mathcal{P}$ from $\mathcal{I}$ and $\mathcal{A}$,
initial GP regression $\hat{g}(x)$\;
Calculate $\check{P}_{\max}((q_0,\delta(q_0,s_0))\models\phi)$ for initial state\;
Calculate $\hat{P}_{\max}((q_0,\delta(q_0,s_0))\models\phi)$ for initial state\;
\While{($\check{P}_{\rm max}<P_{\rm sat}$) \textbf{and} (\rm Count$<$ \rm MaxIterations)}{
\If{$\hat{P}_{\rm max} < P_{\rm sat}$}{
\KwRet{No satisfying control policy exists\;}}
Find nonviolating sub-PIMDP $\mathcal{P}'$ using Algorithm \ref{Sub-Graph Generation}\;
Find MEC to cycle in with corresponding control policy $\pi^*$ using Algorithm \ref{Safe Cycle Selection}\;
\For{NumInnerIterations}{
    Take action $\pi^*(q)$ at current state $q$\;
    Sample $y[k]=x[k+1]-f(x[k])-u[k]$\;
}
Construct GP $\hat{g}(x)$ using collected samples $y[k]$\;
Recalculate transition probability intervals for each state in $\mathcal{P}$\;
Recalculate $\check{P}_{\rm max}((q_0,\delta(q_0,s_0))\models\phi)$ for initial state\;
Recalculate $\hat{P}_{\rm max}((q_0,\delta(q_0,s_0))\models\phi)$ for initial state\;
}
\If{$\check{P}_{\max}\geq P_{\text{sat}}$}{
\KwRet{Control policy $\pi^\dagger=\underset{\alpha\in A(q)}{\mathrm{argmax}}\ \check{P}_{\rm max}((q,s)\models\phi) \ \forall (q,s)\in \mathcal{P}$}\;
}
\end{algorithm}
\end{document}